\documentclass[]{proarticle}
\usepackage{multicol}
\usepackage{graphicx}
\usepackage{subfigure}
\usepackage{amssymb}
\usepackage{epsfig}

\def\bea{\begin{eqnarray}}
\def\eea{\end{eqnarray}}
\begin{document}
\title{The universe evolution in exponential $F(R)$-gravity}
\author{K. Bamba$^1$, A. Lopez-Revelles$^{2,5}$, R. Myrzakulov$^6$, S.~D. Odintsov$^{2,3,4,6}$ and 
L.~Sebastiani$^{5,6}$}
\maketitle
\address{
$^1$KMI,
Nagoya University, Nagoya, Japan\\ 
$^2$ICE/CSIC-IEEC, Bellaterra (Barcelona), Spain
$^3$ ICREA, Barcelona, Spain
$^4$ TSPU, Tomsk, Russia\\
$^5$Dipartimento di Fisica, Universit\`a di Trento, Italia\\
$^6$ Department of General \& Theoretical Physics, Eurasian National University, Astana, Kazakhstan
}
\begin{abstract}
A generic feature of viable exponential $F(R)$-gravity is investigated. An additional modification to stabilize the effective dark energy oscillations during matter era is proposed and applied to two viable models.
An analysis on the future evolution of the universe is performed.
Furthermore, a unified model for early and late-time acceleration is proposed and studied.
\end{abstract}


\begin{multicols}{2}
\section{Introduction}

The current acceleration of our universe is 
supported by various observations and cosmological data \cite{WMAP}.
There exist several descriptions of this acceleration and, among them, the most important
are given by the $\Lambda$CDM Model of
General Relativity (GR) plus Cosmological Constant
and by the modified gravitational theories, 
like $F(R)$-gravity \cite{F(R)}.
Here, we study some aspects and generic feature of viable $F(R)$-gravity models which reproduce the physics of $\Lambda$CDM Model, namely power-law and exponential gravity.
We show 
that 
the behavior of higher derivatives of the Hubble parameter may be 
influenced by large frequency oscillations of effective dark energy during matter era, 
which makes solutions singular and unphysical at a high redshift. 
As a consequence, 
we examine an additional correction term to the models 
in order to remove any instability with keeping the viability properties. 
Some comments about the growth of matter perturbations of corrected models are given and future universe evolution is analyzed.
We also propose a way to describe the early time acceleration of universe
by applying exponential gravity to inflationary cosmology and
a unified description between inflation and current acceleration is presented. 
This work is based on Refs. \cite{OmegaDE,Expgravity}.
We denote the
gravitational constant $8 \pi G_\mathrm{N}$ by 
${\kappa}^2$.

%
\section{Realistic $F(R)$-models in the FLRW universe}\label{sec2}
%
The action describing $F(R)$-gravity is given by 
\begin{equation}
I=\int_{\mathcal{M}} d^4x\sqrt{-g}\left[
\frac{F(R)}{2\kappa^2}+\mathcal{L}^{\mathrm{(matter)}}\right]\,,
\label{action}
\end{equation}
where $F(R)$ is a generic function of the Ricci scalar $R$ only, 
$g$ is the determinant of the metric tensor $g_{\mu\nu}$ and 
${\mathcal{L}}^{\mathrm{(matter)}}$ is the matter Lagrangian.
In realistic models reproducing the standard cosmology of GR with a suitable correction to realize current acceleration 
and/or inflation one has $F(R)=R+f(R)$ and the modification of gravity is encoded in the function $f(R)$. In this case, we may use an effective fluid-like representation of modified gravity and in FLRW space-time described by the metric $ds^{2}=-dt^{2}+a(t)^2 d \mathbf{x}^{2}$, the equations of Motion (EOM) assume the form
\begin{equation}
\rho_{\mathrm{DE}}+\rho_ {\mathrm{m}}= \frac{3}{\kappa^{2}}H^{2}\,, \quad P_{\mathrm{DE}} +P_{\mathrm{m}}= 
-\frac{1}{\kappa^{2}} \left( 2\dot H+3H^{2} \right)\,. \label{EOM}
\end{equation}
Here, $\rho_ {\mathrm{m}}$ and $P_{\mathrm{m}}$ are the energy density and pressure of standard matter and $\rho_{\mathrm{DE}}$ and $P_{\mathrm{DE}}$ are 
the effective dark energy density and pressure given by modified gravity.
In order to study the dynamics of $F(R)$ gravity models, we may introduce the
variable $y_H(z)=\rho_{\mathrm{DE}}/\rho_{\mathrm{m}(0)}$ as a function of the
red shift $z$, where $\rho_{\mathrm{m}(0)}=3\tilde m^2/\kappa^2$ is the matter energy density at the present time, being $\tilde m^2$ a mass scale. From the EOM one derives an equation on the type \cite{Bamba}
\begin{equation}
y_H''(z)+J_1(z)y'_H(z)+J_2(z)
y_H(z)+J_3(z)=0\,,\label{Equazione}
\end{equation}
where $J_1(z)$, $J_2(z)$ and $J_3(z)$ depend on the $F(R)$-model. 
In what follows, we will 
consider viable models representing 
a realistic scenario to account for dark energy, in particular a suitable reparameterization of the Hu-Sawicki model~\cite{HuSaw}, 
\begin{equation}
F(R)=R-2\Lambda\left\{1-\frac{1}{(R/\Lambda)^{4}+1}\right\}\,, 
\label{HS}
\label{model2}
\end{equation}
and a simple example of exponential gravity~\cite{onestep},
\begin{equation}
F(R)=R-2\Lambda\left[1-\mathrm{e}^{-R/\Lambda}\right]\,.
\label{ExpModel}
\end{equation}
This models are very carefully constructed such that 
in the high curvature regime
$F(R\gg\Lambda)=R-2\Lambda$. 
Here, $\Lambda$ plays the role of the Cosmological Constant which decreases in the flat limit ($F(R\rightarrow 0)\rightarrow 0$) where the Minkowski solution is recovered. 
\begin{figure*}[ht!]
\center
\subfigure[]{\includegraphics[width=0.3\textwidth]{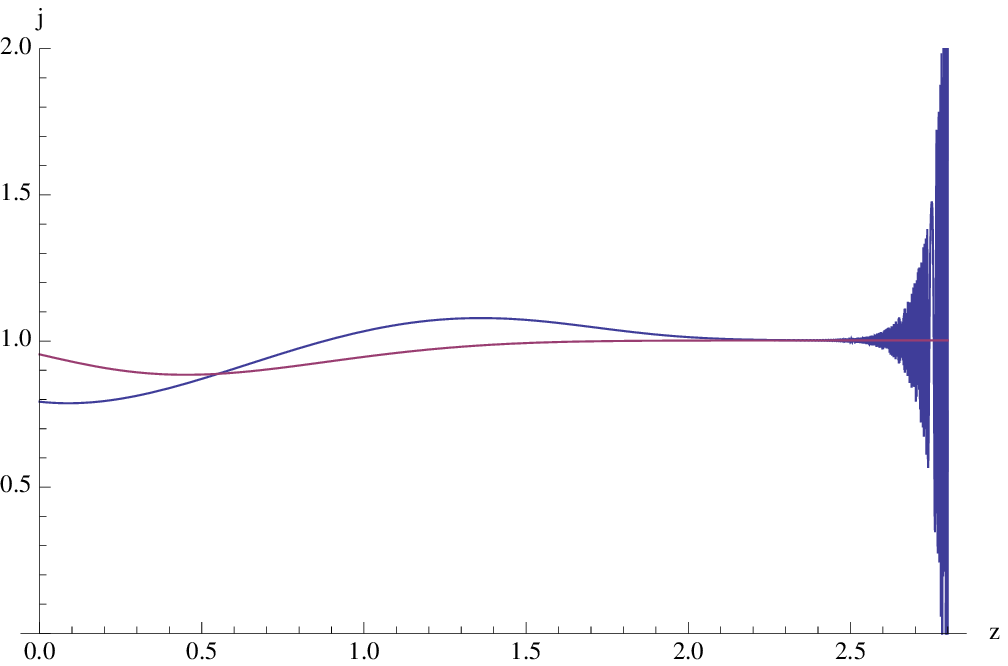}}
\qquad
\subfigure[]{\includegraphics[width=0.3\textwidth]{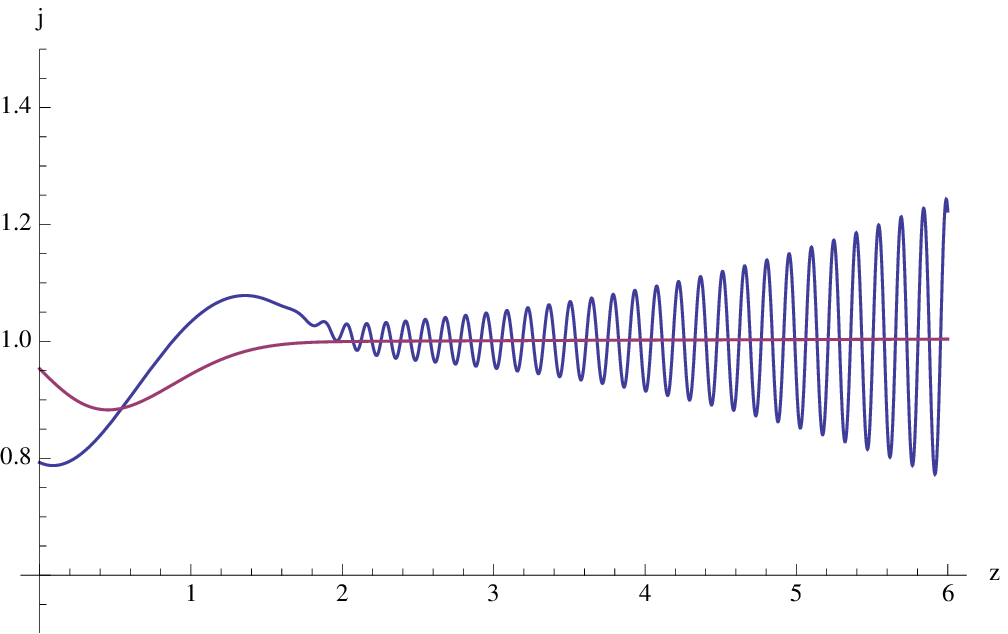}}
\caption{Evolution of the jerk parameter in pure exponential model (a) and in exponential model with correction term (b) overlapped with the ones in $\Lambda$CDM Model.}
\label{Fig1}
\end{figure*} 

\section{DE-oscillations in the matter 
dominated era}\label{sec3}

Since in matter dominated era $R=3\tilde m^2(z+1)^3$, one may locally solve Eq. (\ref{Equazione}) around $z+\delta z$, where $|\delta z|\ll z$. The solution reads
\begin{equation}
y_H(z+\delta z)= a_0+b_0 \delta z + C_0 \mathrm{e}^{{\frac{1}{2(z_0+1)}\left(-\alpha\pm\sqrt{\alpha^2-4\beta}\right)\delta z}}\,,\label{resultmatter}
\end{equation}
where $C_0$ is a generic constant and $a_0$, $b_0$, $\alpha$ and $\beta$ are constants depending on the model. It turns out that the dark energy perturbations remain small around $z$ as soon as the exponential term does not diverge
in expanding universe, when $\delta z<0$.
The explicit conditions $(1-F'(R))F'''(R)/(2F''(R)^2)>-7/2$ and $1/(R F''(R))>12$
are found for stable matter era~\cite{OmegaDE}.
In the case of realistic gravity described by the models (\ref{HS})-(\ref{ExpModel}), where $a_0=\Lambda/(3\tilde m^2)$, it is easy to see that such conditions are well satisfied. Furthermore, since in both of the models $F''(R)\rightarrow 0^+$, 
we understand that 
the discriminant of Eq.~(\ref{resultmatter}) is negative and dark energy oscillates 
with frequency $\mathcal{F}(z)\simeq \left(R*F''(R)\right)^{-1/2}/(z+1)$. 
Since by increasing the curvature $F''(R)$ tends to zero,
for large values of the redshift the dark energy density oscillates 
with a high frequency and also its derivatives become large and may be dominant in some derivatives of the Hubble parameter which approach an effective singularity making unphysical the solution.
Moreover, it may be stated that the closer the model is to  
the $\Lambda$CDM Model (i.e., as much $F''(R)$ is close to zero), 
the bigger the oscillation frequency of dark energy becomes. 
As a consequence, despite the fact that the dynamics of the universe 
depends on the matter and the dark energy density remains very small, 
some divergences in the derivatives of the Hubble parameter can occur, showing a different feature of this kind of models with respect to the case of GR with an undynamic Cosmological Constant. 
Since in the models (\ref{HS})-(\ref{ExpModel})
the approaches to the Cosmological Constant 
are different from each other, 
we may say that 
dark energy oscillations in matter era are 
a generic feature of realistic $F(R)$ gravity, in which 
the cosmological evolution is similar to those in $\Lambda$CDM Model.\\
In order to remove the divergences in the derivatives of the Hubble parameter, 
we can introduce a function $g(R)$ for which the oscillation frequency of 
the dark energy density acquires a constant value 
$\mathcal{F}(z)=1/\sqrt{\delta}$, $\delta>0$,
during matter era,
stabilizing the oscillations 
with the use of a correction term. This term has been derived as
\begin{equation}
g(R) = -\tilde{\gamma}\,\Lambda\left(\frac{R}{3\tilde m^2}\right)^{1/3}\,,
\label{correctionterm}
\end{equation}
where $\tilde{\gamma}=1.702\,\delta$, so that  $\left(R*g''(R)\right)^{-1/2}/(z+1)$
when $R=3\tilde m^2(z+1)^3$.
We explore 
the models (\ref{HS})-(\ref{ExpModel}) 
with adding these correction. 
The effects of the compensating term vanish in the de Sitter epoch, when 
$R=4L$ and the models resemble to a model with an effective 
cosmological constant, provided that $\tilde{\gamma}\ll (\tilde{m}^2/\Lambda)^{1/3}$. 
For example, a reasonable choice is to put $\tilde{\gamma}=1/1000$,  which corresponds 
to the oscillations frequency of the dark energy during matter era $\mathcal{F}=41.255$ with a period $T\simeq 0.152$. 
In order to appreciate the role of the proposed correction, 
let us consider the jerk parameter $j(z)$, which depends on the second derivative of $H(z)$\footnote{The form of the jerk parameter as a function of time
 is $j(t)=a'''(t)/(a(t)H(t))$, $a(t)$ being the scale factor of the universe.}, for the exponential model (\ref{ExpModel}), whose numerical extrapolation\footnote{We have used 
Mathematica 7 \textcopyright.} is plotted and overlapped with its behavior in $\Lambda$CDM Model on the left side of Fig. \ref{Fig1}. Despite the fact that the dark energy density $\rho_{\mathrm{DE}}\simeq\Lambda/\kappa^2$ is negligible at high red shift, the effects of its oscillations become relevant in the jerk parameter, which 
grows up with an oscillatory behavior and diverges at $z\simeq 2.8$. On the right side of the figure, we plot the graphic of the jerk parameter for the same model with the addition of the correction term $g(R)$ previously discussed. 
Also in this case the jerk parameter oscillates with the same frequency of the dark energy, but here such oscillations have a constant frequency and do not diverge at high red shift.\\ 
It is also important to stress that
the term added to 
stabilize the dark energy oscillations in the matter dominated epoch 
does not cause any problem to the good proprieties of the models (\ref{HS})-(\ref{ExpModel}), which continue to satisfy all the cosmological and local gravity 
constraints.
It has been demonstrated in several papers that in order to evade the solar system tests ~\cite{Solarsystem} or avoid large corrections to Newton Law at large cosmological scales~\cite{matterinstabilityplanet}, a modified gravity theory must satisfy the following relation
\begin{equation}
\frac{1}{3R}\left(\frac{F'(R)}{F''(R)}-R\right)\gg 1\,,
\end{equation}
when $R$  assumes the typical curvature value of the constant background around a matter source. For the models~(\ref{HS})-(\ref{ExpModel}) without 
correction terms, this condition is always well satisfied, due to the fact that $F''(R\gg\Lambda)\simeq 0^+$.
Next, if we consider the addition of $g(R)$, whose second derivative is dominant, the left side of the letter equation reads
$3^{4/3}\tilde m^2 /(2\Lambda\tilde{\gamma})\left(R/\tilde m^2\right)^{2/3}$.
Despite the fact that this quantity is smaller than the one obtained for the pure models (\ref{HS})-(\ref{ExpModel}), 
it still remains sufficiently large and the correction to the Newton Law is 
very small. 
For example, at the typical value of the curvature in the solar system, namely
$R\simeq 10^{-61}$eV$^2$,
it corresponds to $2\times 10^{6}$, which is very large effectively. 
Concerning the matter instability~\cite{Faraoni}, this might also occur when the curvature is 
rather large, as on a planet ($R\simeq 10^{-38}$eV$^2$), 
as compared with the average curvature of the 
universe today ($ R\simeq 10^{-66}$eV$^2$). 
This instability is avoided as soon as $1/(R F''(R))$ is much larger than one on the planet surfaces. 
For our correction term $g(R)$, this quantity evaluated at $R\simeq 10^{-38}$eV$^2$ corresponds to $4\times 10^{21}$
and also in this case we do not have any particular problem.

\section{Growth index}\label{sec4}

After the discovery of cosmic acceleration, a lot of 
theories for the dark energy have been proposed, and it has become very important
to find a way to discriminating 
among these theories, most of them 
exhibit the same expansion history. 
In this context, the characterization of growth of the matter density perturbations is very significant in modified gravity, since models with the same cosmological evolution have a different cosmic growth history. 
In order to execute it, the so-called growth index $\gamma(z)$ is useful~\cite{Linder:2005in}.
Under the subhorizon approximation, 
the matter density perturbation 
$\delta = \delta \rho_\mathrm{m}/\rho_\mathrm{m}$ 
satisfies the equation
$\ddot \delta \, + \, 2 H \dot \delta \, - \, 4 \pi G_\mathrm{eff}(a,k) \rho_\mathrm{m} \delta = 0$ with $k$ being the comoving wavenumber and $G_\mathrm{eff}(a,k)$ being 
the effective gravitational ``constant'' depending on the model and expressed as a function of $k$ and the scale factor a, such that $k^2/a^2\gg H$. The equation for matter perturbation has an equivalent description in terms of the growth rate $f_\mathrm{g} \equiv d \ln{\delta}/d \ln{a}$. The growth index $\gamma(z)$ is defined as  is defined as the quantity satisfying the relation
$f_\mathrm{g}(z) = \Omega_\mathrm{m}(z)^{\gamma(z)}$
with $\Omega_\mathrm{m}(z) = 8 \pi G \rho_m/3 H^2$ being 
the matter density parameter. 
The growth index cannot be 
observed directly, but it can be determined from the observational data of both the growth factor and the matter density parameter at the same redshift and this is the reason for which in the last period its calculation in different theories has become of a great interest.
In Ref.~\cite{Expgravity} various parameterizations for the growth index $\gamma(z)$ have been studied for the models (\ref{HS})-(\ref{ExpModel}) with correction term $g(R)$.
It has been found that the Ansatz which better fits the (numerical) solution of the growth rate  
for the considered models is given by $\gamma = \gamma_0 + \gamma_1 z$, where $\gamma_{0,1}$ are constants. 

\section{Future universe evolution}\label{sec5}

The confrontation of the models~(\ref{HS})-(\ref{ExpModel}) with SNIa, BAO, CMB radiation 
and gravitational lensing has been executed in the past in several 
works and all the numerical analysis have demonstrated that they are consistent 
with the last observational data coming from our universe. In particular, they reproduce the correct amount of dark energy today $\Omega_\mathrm{DE} = 0.721\pm 0.015$~\cite{Bamba, works}. 
Moreover, the addition of correction term (\ref{correctionterm}) does not have any influence at the present day range of detection.
Starting from Eq.~(\ref{Equazione}), we are able to study perturbations around 
the de Sitter solution $R_{\mathrm{dS}}=4\Lambda$ of our universe today in the considered models.
Performing the variation with respect to $y_H(z)=y_0+y_1(z)$ with $y_0=R_{\mathrm{dS}}/12\tilde{m}^2$ and $|y_1(z)|\ll 1$ and assuming the contributions of radiation and matter to be much smaller 
than $y_0$, one finds the following solution for $y_1(z)$,
\begin{equation}
y_1(z) = C_0(z+1)^{\frac{1}{2}\left(3\pm\sqrt{9-4\beta}\right)} + \frac{4\zeta}{\beta}(z+1)^3\,,
\label{result}
\end{equation}
where $C_0$ is a generic constant and $\zeta=1+(1-F'(R))/(R F''(R))$ and $\beta=-4+4F'(R)/(RF''(R))$, the both evaluated at $R=R_{\mathrm{dS}}$. The de Sitter solution is stable provided by condition $F'(R_\mathrm{dS})/((R_\mathrm{dS})F''(R_\mathrm{dS}))>1$, which is a well know result. 
It has also been demonstrated that since in realistic $F(R)$-gravity models for the de Sitter universe 
$0<R_{\mathrm{dS}}F''(R_{\mathrm{dS}})\ll 1$, one has  
$F'(R_{\mathrm{dS}})/(R_{\mathrm{dS}}F''(R_{\mathrm{dS}}))>25/16$, giving the negative discriminant of Eq.~(\ref{result}) and an oscillatory behavior to the dark energy density during this phase. 
Thus, in this case 
the dark energy Equation of State (EoS) parameter $\omega_\mathrm{DE}\equiv P_\mathrm{DE}/\rho_\mathrm{DE}$ oscillates infinitely often around the line of the phantom divide 
$\omega_\mathrm{DE}=-1$~\cite{Staro}. 
These models possess one crossing in the recent past, after the end of the matter dominated era, and infinite crossings in the future, but the amplitude of such crossings decreases as $(z+1)^{3/2}$ and it does not cause any serious problem to the accuracy of the cosmological evolution during the de Sitter epoch which is in general the final 
attractor of the system. 
However, since the existence of a phantom phase can give some undesirable effects 
such as 
the possibility to have the Big Rip~\cite{Caldwell} or Big Rip scenario with 
the disintegration of bound structures~\cite{Rip}, it is important 
to observe that in the models~(\ref{HS})-(\ref{ExpModel}) with the adding modification (\ref{correctionterm}), the effective EoS parameter of the universe 
\begin{equation}
\omega_{\mathrm{eff}}\equiv\frac{\rho_{\mathrm{DE}}+\rho_{\mathrm{m}}}{P_{\mathrm{DE}}+P_{\mathrm{m}}}=-1+\frac{2(z+1)}{3H(z)}\frac{dH(z)}{dz}\,, 
\end{equation}
never crosses the phantom divide line in the past, and that only in the very far future, when $z$ is very close to $-1$, 
it coincides with $\omega_{\mathrm{DE}}$ and the crossings occur. 
Let us consider for example the exponential case (\ref{ExpModel}) plus term (\ref{correctionterm}).
The numerical evaluation of Eq.~(\ref{Equazione}) leads to 
Hubble parameter
\begin{equation}
H(z)=\sqrt{\tilde m^2\left[y_H(z)+(z+1)^3\right]}\,,
\end{equation}
and therefore $\omega_{\mathrm{eff}}(z)$. 
We depict the corresponding cosmological evolution 
of $\omega_{\mathrm{eff}}(z)$ for $-1<z<2$ 
in Fig.~2.
We can see that 
$\omega_{\mathrm{eff}}(z)$ starts from zero in the matter dominated 
era and asymptotically approaches -1 without any appreciable deviation. 
Only when $z\simeq -0.90$,  $\omega_{\mathrm{\mathrm{eff}}}(z)$ crosses the line of phantom divide, how it is shown on the right.
By considering the scale factor $a(t)=\exp \left(H_0 t\right)$, 
where $H_0\simeq 6.3\times 10^{-34}$eV$^{-1}$ is the Hubble parameter of 
the de Sitter universe, we may conclude that the crossings appear at $10^{26}$ years. 

\includegraphics[width=0.3\textwidth]{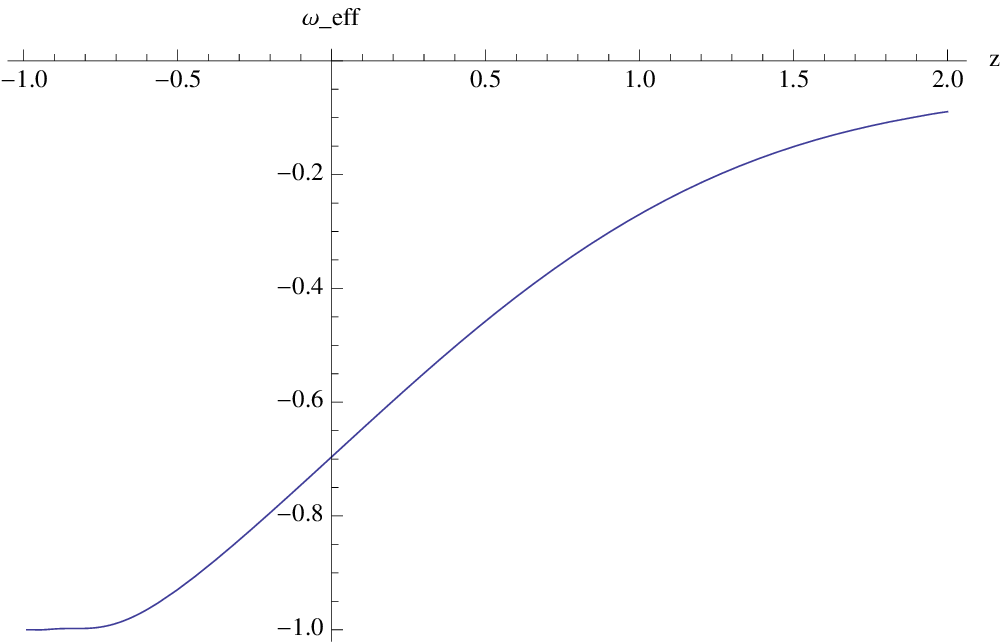}
\includegraphics[width=0.15\textwidth]{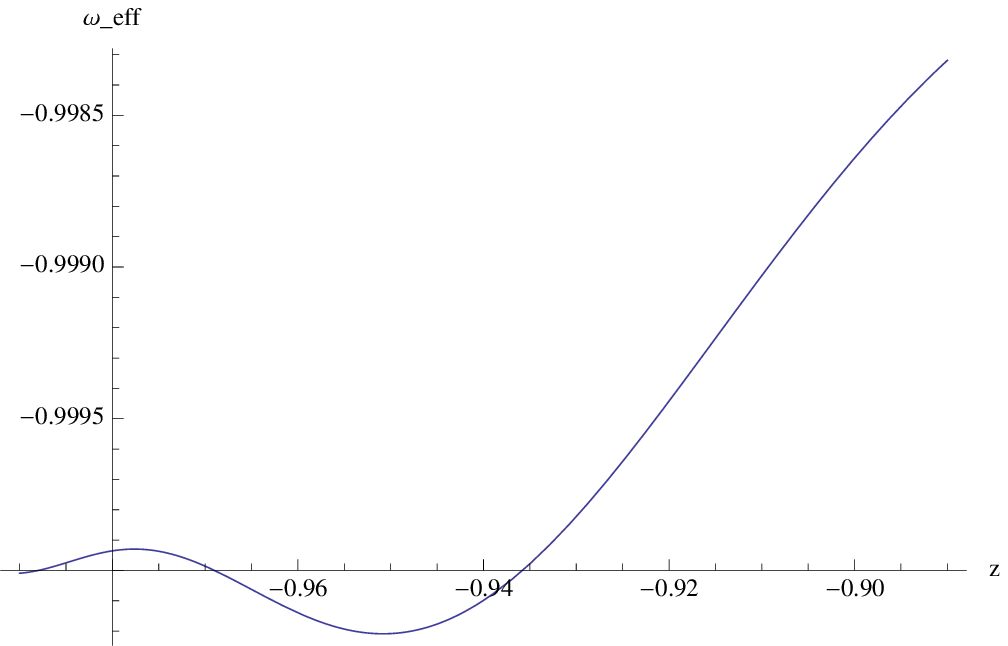}
\\
Figure 2: Effective EoS parameter for $-1<z<2$ in exponential model. On the right, the region $-1<z<-0.9$ is stressed.

\section{Exponential gravity for early and late-time cosmic acceleration}\label{sec5}

One of the most important goals for modified gravity theories is that 
all the processes in the expansion history of the universe are 
realized, from inflation to
the late cosmic acceleration epoch, 
without invoking the presence of dark components.
Models of the type~(\ref{ExpModel}) may be combined in a natural way 
to obtain the phenomenological description of the inflationary epoch, mimicking a smooth version of $F(R)=R-2\Lambda \left[1-\mathrm{e}^{-R/\Lambda}\right]-\Lambda_{\rm i} \,\theta(R-R_\mathrm{i})$, where $R_i$ is the transition scalar curvature for which a suitable cosmological constant $\Lambda_i$ switches on producing inflation. 
Following the first proposal of Ref.~\cite{twostep}, 
we consider the form of $F(R)$ with a natural possibility of 
a unified description of our universe 
\begin{eqnarray}
F(R)&=&R-2\Lambda\left(1-\mathrm{e}^{-\frac{R}{\Lambda}}\right)
-\Lambda_\mathrm{i}\left[1-\mathrm{e}^{-\left(\frac{R}{R_\mathrm{i}}\right)^n}\right]\nonumber\\
&&+\bar{\gamma} \left(\frac{1}{\tilde R_\mathrm{i}^{\alpha-1}}\right)R^\alpha\,, 
\label{total}
\end{eqnarray}
where $R_\mathrm{i}$ and $\Lambda_\mathrm{i}$ are the typical values of 
transition curvature and expected cosmological constant during inflation, 
respectively, and $n$ is a natural number larger than unity 
(here, we do not write the correction term for the stability of 
oscillations in the matter dominated era). 
In Eq.~(\ref{total}), 
the last term 
$\bar{\gamma}(1/\tilde R_\mathrm{i}^{\alpha-1}) R^\alpha$, 
where $\bar{\gamma}$ 
is a positive dimensional constant and $\alpha$ is a real number, 
works at the inflation scale $\tilde R_\mathrm{i}$ and 
is actually necessary in order to realize an exit from inflation. 
An accurate analysis of this model shows that if $\alpha>1$ and $n>1$, we avoid the effects 
of inflation at small curvatures and it does not influence the stability of 
the matter dominated era. 
The de Sitter point $R_{\mathrm{dS}}$ which describes the early time acceleration exists if $\tilde R_\mathrm{i}=R_{\mathrm{dS}}$ and 
it reads as $R_{\mathrm{dS}}=2/(\bar{\gamma}(2-\alpha)+1)$ under the condition $(R_{\mathrm{dS}}/R_\mathrm{i})^n\gg 1$. 
Thus, the inflation is unstable if $\alpha\bar\gamma(\alpha-2)>1$. 
We may evaluate the characteristic number of $e$-folds during inflation 
$N=\log [(z_\mathrm{i}+1)/(z_\mathrm{e}+1)]$,
where $z_\mathrm{i}$ and $z_\mathrm{e}$ are the redshifts at the beginning and at the end of early time cosmic acceleration. 
Given a small cosmological perturbation $y_1(z_\mathrm{i})$ at the redshift 
$z_\mathrm{i}$ to the effective dark energy density of inflation $y_0=R_{\mathrm{dS}}/(12\tilde m^2)$ , so that $y(z)=y_0+y_1(z)$, we have from Eq.~(\ref{Equazione}) avoiding 
the matter contribution
\begin{equation}
y_1(z_\mathrm{i})=C_0(z_\mathrm{i}+1)^x\,, 
\label{pertinfl}
\end{equation}
with
\begin{equation}
x=\frac{1}{2}\left(3-\sqrt{25-\frac{16F'(R_{\mathrm{dS}})}{R_{\mathrm{dS}}F''(\mathrm{R_{dS}})}}\right)\,.
\label{x}
\end{equation}
Here, $x<0$ if the de Sitter point is unstable. As a consequence, the perturbation $y_1(z)$ grows up in expanding universe as
$y_1(z)=y_1(z_\mathrm{i})\left[(z+1)/(z_\mathrm{i}+1)\right]^x$.
When $y_1(z)$ is on the same order of the effective modified gravity energy density $y_0(z)$ of 
the de Sitter solution describing inflation, the model exits from 
inflation. In this way, we can estimate the number of $e$-folds during inflation 
as
\begin{equation}
N\simeq \frac{1}{x}\log\left(\frac{y_1(z_\mathrm{i})}{y_0}\right)\,.
\label{Nfolding}
\end{equation}
A value demanded in most inflationary scenarios is at least 
$N = 50$--$60$. A classical perturbation on the (vacuum) de Sitter solution may be given by 
the presence of ultrarelativistic matter in the early universe. Thus, we can predict the numbers of 
$e$-folds for the model (\ref{total}) by computing $x$ in Eq. (\ref{x}), which depends on model parameters. 
Let us consider an example by supposing that ultrarelativistic matter/radiation is
$10^6$ times smaller than 
that of effective modified gravity energy density($\sim \Lambda_{\mathrm{i}}$) at the beginning of inflation.
By choosing the viable model parameters $n=4$, $R_i=2\Lambda_i$, $\bar{\gamma}=1$ and $\alpha=8/3$, such that the curvature at the inflation results to be $R_{\mathrm{dS}}=6\Lambda_{\mathrm{i}}$, one has $x=-0.218$ and, by using Eq. (\ref{Nfolding}), we obtain $N\simeq 64$ and $z_{\mathrm{e}}=e^{-64}(z_{\mathrm{i}}+1)-1$. The numerical extrapolation of Eq. (\ref{Equazione}) gives 
us $y_H(z_\mathrm{e})=0.88y_H(z_\mathrm{i})$ and $R(z_\mathrm{e})=0.853R_{\mathrm{dS}}$, confirming that the effective modified gravity energy density and the curvature decrease at the end of inflation. After that, it is reasonable to expect that 
the physical processes described by the $\Lambda$CDM Model are reproduced by the first term of modification to gravity. 

\section{Conclusions}

In this work, we have analyzed some feature of realistic $F(R)$-gravity mimicing the physics of $\Lambda$CDM Model. Two viable models have been considered.
We have seen that since the corrections to GR at small curvatures 
may lead to undesired effects at high curvatures, 
some adding modifications are required. In particular,
we have reconstructed a compensating term 
in order to stabilize the oscillations of the effective dark energy at high red shifts with retaining the viability proprieties. Furthermore, we have briefly discussed about the growth index and an investigation on the cosmological evolution of the 
universe has been executed showing that the effective crossing of the phantom divide, 
which characterizes the de Sitter epoch, occurs in the very far future. 
We also have considered the inflationary cosmology in exponential gravity. It has been derived the correlation between the $e$-folds during 
inflation and the model parameters. Since at the end of inflation the effective modified gravity energy density and the curvature decrease, we may conclude that it is possible to acquire 
a gravitational alternative scenario for a unified description of inflation in 
the early universe with the late-time cosmic acceleration.


\end{multicols}
\end{document}